\documentclass[a4]{article}
\begin{document}

\author{M. Matlak$^{a}\thanks{%
Corresponding author, e-mail: matlak@us.edu.pl}$, B. Grabiec$^{b }$,
S. Krawiec$^{a}$ \\
$^{a}$Institute of Physics, University of Silesia, Uniwersytecka 4, \\
40-007 Katowice, Poland\\
$^{b}$Institute of Physics, University of Zielona G\'{o}ra,\\
Prof. Z. Szafrana 4a, 65-516 Zielona G\'ora, Poland}
\title{Electronic Correlations within Fermionic Lattice Models }
\maketitle

\begin{abstract}
We investigate two-site electronic correlations within generalized Hubbard
model, which incorporates the conventional Hubbard model
(parameters: $t$ (hopping between nearest neighbours), $U$ (Coulomb
repulsion (attraction)) supplemented by the intersite Coulomb interactions
(parameters: $J^{(1)}$(parallel spins), $J^{(2)}$ (antiparellel spins)) and the hopping of the intrasite Cooper pairs (parameter: $V$). As a
first step we find the eigenvalues $E_{\alpha} $ and eigenvectors $|E_{\alpha} \rangle $ of the dimer and we represent each partial Hamiltonian $E_{\alpha} |E_{\alpha} \rangle \langle E_{\alpha} |$ ($\alpha =1,2,..,16$) in the second
quantization with the use of the Hubbard and spin operators. Each dimer energy level possesses its own Hamiltonian describing different two-site interactions which can
be active only in the case when the level will be occupied by the electrons.
A typical feature is the appearence of two generalized $t-J$ interactions
ascribed to two different energy levels which do not vanish even for $%
U=J^{(1)}=J^{(2)}=V=0$ and their coupling constants are equal to $\pm t$ in
this case. In the large $U$ - limit for $J^{(1)}=J^{(2)}=V=0$ there is only
one $t-J$ interaction with coupling constant equal to $4t^2/ \left|
U\right|$ as in the case of a real lattice. The competition between ferromagnetism, antiferromagnetism and superconductivity (intrasite and intersite pairings) is also a typical feature of the model because it persists in the case $U=J^{(1)}=J^{(2)}=V=0$ and $t\neq 0$. The same types of the electronic, competitive interactions are scattered between different energy levels and therefore their
thermodynamical activities are dependent on the occupation of these levels.
It qualitatively explains the origin of the phase diagram of the
model. We consider also a real lattice as a set of interacting dimers to show that the competition between magnetism and superconductivity seems to be universal for fermonic lattice models. 

\end{abstract}

\section{Introduction}
Model Hamiltonians which serve to describe electronic subsystems of solids
are formulated on the quantum mechanical basis which takes into account
interaction processes between electrons, most of them basing on the Hubbard
model and its generalizations (cf e.g. Refs [1]-[96]). The ground state
properties of such models has been investigated in Refs [85]-[92]. Each
model Hamiltonian of this type desribes, however, as a rule many unknown,
competitive, electronic correlations (two-site, three-site, etc.) which are
normally invisible in the original model but the knowledge about their
existence is very important because just these correlations determine also
an area of physics where a given model Hamiltonian can really be applied. To
find the electronic correlation within a given model Hamiltonian $H$ let us
assume that we can exactly solve the Schr\"{o}dinger equation $H|E_\alpha
\rangle =E_\alpha |E_\alpha \rangle $. Having to our disposal the calculated
eigenvalues $E_\alpha $ and eigenvectors $|E_\alpha \rangle $ we can use the
equivalent form of the given Hamiltonian $H=\sum\limits_\alpha H_\alpha $
where the set of commuting partial Hamiltonians $H_\alpha =E_\alpha
|E_\alpha \rangle \langle E_\alpha |$ we can represent in the second
quantization for each energy level separately introducing here Hubbard and
spin operators. In this way to each energy level of the system $E_\alpha $ a
partial Hamiltonian $H_\alpha $ can be ascribed. Each part $H_\alpha $
contains many important, competitive interactions, active in the case when
the level will be occupied by the electrons. This simple idea cannot be
unfortunately applied in a general case because we cannot exactly solve the
mentioned Schr\"{o}dinger equation. It is, however, possible to do it
exactly and analytically in the case of a dimer described by the generalized
Hubbard model (see Sec. 2) to show that two-site electronic correlations
resulting from this approach describe the competition between magnetism and
superconductivity. We show that this competition is an universal feature of
all electronic lattice models containing hopping term and it takes also
place in the case of a real lattice. This result seems to be very important
also in the case of quantum dots and nanostructures (cf e.g. Refs
[97]-[100]).

\section{ The Model}
The generalized one-band Hubbard model belongs to a class of fermionic
lattice models widely used in the solid state physics. This model has been
primarily applied to explain magnetic and transport properties of transition
metals, their compounds and alloys, including also insulator-metal
transitions (cf e.g. Refs [1]-[16] and original papers cited therein). After
further generalizations the model has also been applied to fluctuating
valence systems and heavy fermions (Anderson-like models, Refs [17]-[18]
(see also Ref. [13] for a review)), liquide $^3He$ (see e.g. Refs [19]-[21])
and fullerenes (cf e.g. Refs [22]-[24]). A special attention has been,
however, paid in recent decades to the theory of high-$T_C$
superconductivity (cf Refs [25]-[58]) where the extended Hubbard model
(negative $U$ model) has widely been used (see e.g. Ref. [33] for a review).
Another interesting model also used in the context of superconductivity is
the KPPK model, formulated in Refs [59], [60] (see also e.g. Refs
[61]-[64]). In the present paper we consider the extended Hubbard model
supplemented by the hopping of the intrasite Cooper pairs (KPPK
interaction). The Hamiltonian of this model has the form 
\begin{equation}
\begin{array}{ll}
H= & \sum\limits_{i\neq j,\sigma }t_{i,j}c_{i,\sigma }^{+}c_{j,\sigma
}+U\sum\limits_in_{i,\uparrow }n_{i,\downarrow }+\frac 12\sum\limits_{i\neq
j,\sigma }J_{ij}^{(1)}n_{i,\sigma }n_{j,\sigma } \\ 
& +\frac 12\sum\limits_{i\neq j,\sigma }J_{ij}^{(2)}n_{i,\sigma
}n_{j,-\sigma }-\sum\limits_{i\neq j}V_{i,j}c_{i,\uparrow
}^{+}c_{i,\downarrow }^{+}c_{j,\downarrow }c_{j,\uparrow }.
\end{array}
\end{equation}
The indices ($i,j$) enumerate the lattice points $(\mathbf{R_i,R_j)}$, $%
t_{i,j}$ is the hopping integral, $U$ denotes the effective intrasite
Coloumb interaction, $J^{(1)}$ and $J^{(2)}$ (generally, not necessary
equal) describe the effective intersite interactions, all of them resulting
from the original intrasite and intersite Coulomb repulsion which can be
modified by polaronic effects (see e.g. Ref. [33] for details) and therefore 
$U,$ $J^{(1,2)}$ can be treated here as positive or negative parameters. The
last term in (1) is responsible for the transport of the intrasite Cooper
pairs (Refs [59], [60]) with the coupling constant $V$. The model (1) cannot
be solved exactly in a general case. We can, however, consider a special but
nontrivial case of two interacting ions (a dimer problem) which posseses
exact, analytical solution. Thus, let us start with the dimer Hamiltonian,
resulting from the expression (1). It has the form

\begin{equation}
\begin{tabular}{ll}
$H_D=$ & $-t\sum\limits_{\sigma} (c_{1,\sigma }^{+}c_{2,\sigma }+c_{2,\sigma
}^{+}c_{1,\sigma })+U(n_{1,\uparrow }n_{1,\downarrow }+n_{2,\uparrow
}n_{2,\downarrow })$ \\ 
& $+J^{(1)}\sum\limits_{\sigma} n_{1,\sigma }n_{2,\sigma
}+J^{(2)}\sum\limits_{\sigma} n_{1,\sigma }n_{2,-\sigma }$ \\ 
& $-V(c_{1,\uparrow }^{+}c_{1,\downarrow }^{+}c_{2,\downarrow }c_{2,\uparrow
}+c_{2,\uparrow }^{+}c_{2,\downarrow }^{+}c_{1,\downarrow }c_{1,\uparrow })$%
\end{tabular}
\end{equation}

where $t_{1,2}=t_{2,1}=-t$, $J_{1,2}^{(1,2)}=J_{2,1}^{(1,2)}=J^{(1,2)}$ and $%
V_{1,2}=V_{2,1}=V$. We start from the Fock's basis $|n_{1,\uparrow
},n_{1,\downarrow };n_{2,\uparrow },n_{2,\downarrow }\rangle $ $(n_{i,\sigma
}=0,1;$ $i=1,2$; $\sigma =\uparrow ,\downarrow )$ and we find the exact
solution of the dimer eigenvalue problem ($H_D|E_{\alpha} \rangle
=E_{\alpha} |E_{\alpha} \rangle $):

\[
\begin{array}{ll}
E_1=0; & |E_1\rangle =|0,0;0,0\rangle ,
\end{array}
\]

\[
\begin{array}{ll}
E_2=-t; & |E_2\rangle =\frac 1{\sqrt{2}}(|1,0;0,0\rangle +|0,0;1,0\rangle ),
\\ 
E_3=t; & |E_3\rangle =\frac 1{\sqrt{2}}(|1,0;0,0\rangle -|0,0;1,0\rangle ),
\\ 
E_4=-t; & |E_4\rangle =\frac 1{\sqrt{2}}(|0,1;0,0\rangle +|0,0;0,1\rangle ),
\\ 
E_5=t; & |E_5\rangle =\frac 1{\sqrt{2}}(|0,1;0,0\rangle -|0,0;0,1\rangle ),
\end{array}
\]

\[
\begin{array}{ll}
E_6=J^{(2)}; & |E_6\rangle =\frac 1{\sqrt{2}}(|1,0;0,1\rangle
+|0,1;1,0\rangle ), \\ 
E_7=U+V; & |E_7\rangle =\frac 1{\sqrt{2}}(|1,1;0,0\rangle -|0,0;1,1\rangle ),
\\ 
E_8=C+\frac{U-V+J^{(2)}}2; & |E_8\rangle =a_{+}(|1,1;0,0\rangle
+|0,0;1,1\rangle )-a_{-}(|1,0;0,1\rangle -|0,1;1,0\rangle ), \\ 
E_9=-C+\frac{U-V+J^{(2)}}2; & |E_9\rangle =a_{-}(|1,1;0,0\rangle
+|0,0;1,1\rangle )+a_{+}(|1,0;0,1\rangle -|0,1;1,0\rangle ), \\ 
E_{10}=J^{(1)}; & |E_{10}\rangle =|1,0;1,0\rangle , \\ 
E_{11}=J^{(1)}; & |E_{11}\rangle =|0,1;0,1\rangle ,
\end{array}
\]

\begin{equation}
\begin{array}{ll}
E_{12}=t+U+J^{(1)}+J^{(2)}; & |E_{12}\rangle =\frac 1{\sqrt{2}%
}(|0,1;1,1\rangle +|1,1;0,1\rangle ), \\ 
E_{13}=-t+U+J^{(1)}+J^{(2)}; & |E_{13}\rangle =\frac 1{\sqrt{2}%
}(|0,1;1,1\rangle -|1,1;0,1\rangle ), \\ 
E_{14}=t+U+J^{(1)}+J^{(2)}; & |E_{14}\rangle =\frac 1{\sqrt{2}%
}(|1,0;1,1\rangle +|1,1;1,0\rangle ), \\ 
E_{15}=-t+U+J^{(1)}+J^{(2)}; & |E_{15}\rangle =\frac 1{\sqrt{2}%
}(|1,0;1,1\rangle -|1,1;1,0\rangle ),
\end{array}
\end{equation}

\[
\begin{array}{ll}
E_{16}=2(U+J^{(1)}+J^{(2)}); & |E_{16}\rangle =|1,1;1,1\rangle
\end{array}
\]
where

\begin{equation}
\begin{array}{ll}
C=\sqrt{{\left( \frac{U-V-J^{(2)}}2\right) }^2+4t^2}, & a_{\pm }=\frac 12%
\sqrt{1\pm \frac{(U-V-J^{(2)})}{2C}}.
\end{array}
\end{equation}

In the following we apply Hubbard and spin operators (cf e.g. Ref. [1])

\begin{equation}
\begin{array}{ll}
a_{i,\sigma }=c_{i,\sigma }(1-n_{i,-\sigma }), & b_{i,\sigma }=c_{i,\sigma
}n_{i,-\sigma },
\end{array}
\end{equation}

\begin{equation}
\begin{tabular}{ll}
$S_i^z=\frac 12(n_{i,\uparrow }^a-n_{i,\downarrow }^a),$ & $n_{i,\sigma
}^a=a_{i,\sigma }^{+}a_{i,\sigma },$ \\ 
&  \\ 
$S_i^{+}=c_{i,\uparrow }^{+}c_{i,\downarrow }=a_{i,\uparrow
}^{+}a_{i,\downarrow },$ & $S_i^{-}=c_{i,\downarrow }^{+}c_{i,\uparrow
}=a_{i,\downarrow }^{+}a_{i,\uparrow }$%
\end{tabular}
\end{equation}
and we use the equivalent expression for the dimer Hamiltonian (2) 
\begin{equation}
H_D=\sum\limits_{\alpha =1}^{16}E_{\alpha} P_{\alpha}
\end{equation}
where $P_{\alpha} =|E_{\alpha} \rangle \langle E_{\alpha} |$. Each product $%
E_{\alpha} P_{\alpha} $ in the formula (7) where we insert $E_{\alpha} $ and 
$|E_{\alpha} \rangle $ from the formulae (3) can be rewritten in the second
quantization with the use of the Hubbard and spin operators (5) and (6). It
is convenient to collect all the terms which correspond to the same energy
level (as e.g. $E_4=E_2$, $E_5=E_3$, $E_{11}=E_{10}$, $E_{14}=E_{12}$, $%
E_{15}=E_{13}$) and the same number of particles $N$. Using (7) we can split 
$H_D$ into 10 terms, corresponding to 10 different dimer energy levels (see
(3)) and belonging to different subspaces of the total number of particles $%
N $. We obtain

\begin{equation}
H_D=\sum_{i=1}^{10}H_D^{(i)}
\end{equation}

where

\begin{eqnarray}
H_D^{(1)} &=&E_2P_2+E_4P_4=-\frac t2[n_1^a(1-n_2^a-\frac{n_2^b}%
2)+n_2^a(1-n_1^a-\frac{n_1^b}2)]  \nonumber \\
&&-\frac t2\sum_{\sigma} [a_{1,\sigma }^{+}a_{2,\sigma }+a_{2,\sigma
}^{+}a_{1,\sigma }], \\
H_D^{(2)} &=&E_3P_3+E_5P_5=\frac t2[n_1^a(1-n_2^a-\frac{n_2^b}%
2)+n_2^a(1-n_1^a-\frac{n_1^b}2)]  \nonumber \\
&&-\frac t2\sum_{\sigma} [a_{1,\sigma }^{+}a_{2,\sigma }+a_{2,\sigma
}^{+}a_{1,\sigma }],
\end{eqnarray}

\begin{equation}
H_D^{(3)}=E_6P_6=-J^{(2)}[{S_1}^z\cdot {S_2}^z-\frac{n_1^an_2^a}4]+\frac{%
J^{(2)}}2\left( {S_1}^{+}\cdot {S_2}^{-}+{S_1}^{-}\cdot {S_2}^{+}\right)
\end{equation}

\begin{equation}
\begin{array}{ll}
H_D^{(4)}=E_7P_7= & \frac{\left( U+V\right) }4[n_1^b(1-n_2^a-\frac{n_2^b}%
2)+n_2^b(1-n_1^a-\frac{n_1^b}2)] \\ 
&  \\ 
& -\frac{\left( U+V\right) }2[d_1^{+}d_2+d_2^{+}d_1],
\end{array}
\end{equation}

\begin{eqnarray}
H_D^{(5)} &=&E_8P_8=\left\{ -\frac{J^{(2)}}2+[\frac{J^{(2)}(U-V-J^{(2)})}{4C}%
-\frac{2t^2}C]\right\} [{\mathbf{\overrightarrow{S_1}\cdot \overrightarrow{%
S_2}}}-\frac{n_1^an_2^a}4]  \nonumber \\
&&+\left[ \frac{\left( U-V\right) }4\left( 1+\frac{U-V-J^{(2)}}{2C}\right) +%
\frac{t^2}C\right] [d_1^{+}d_2+d_2^{+}d_1]  \nonumber \\
&&+\left[ \frac{\left( U-V\right) }8\left( 1+\frac{U-V-J^{(2)}}{2C}\right) +%
\frac{t^2}{2C}\right] [n_1^b(1-n_2^a-\frac{n_2^b}2)+n_2^b(1-n_1^a-\frac{n_1^b%
}2)]  \nonumber \\
&&-\frac t2\left[ 1+\frac{U-V+J^{(2)})}{2C}\right] \sum_{\sigma}
\sum_{i=1}^2[a_{i,\sigma }^{+}b_{\overline{i},\sigma }+b_{i,\sigma }^{+}a_{%
\overline{i},\sigma }],
\end{eqnarray}

\begin{eqnarray}
H_D^{(6)} &=&E_9P_9=\left\{ -\frac{J^{(2)}}2-[\frac{J^{(2)}(U-V-J^{(2)})}{4C}%
-\frac{2t^2}C]\right\} [{\mathbf{\overrightarrow{S_1}\cdot \overrightarrow{%
S_2}}}-\frac{n_1^an_2^a}4]  \nonumber \\
&&+\left[ \frac{\left( U-V\right) }4\left( 1-\frac{U-V-J^{(2)}}{2C}\right) -%
\frac{t^2}C\right] [d_1^{+}d_2+d_2^{+}d_1]  \nonumber \\
&&+\left[ \frac{\left( U-V\right) }8\left( 1-\frac{U-V-J^{(2)}}{2C}\right) -%
\frac{t^2}{2C}\right] [n_1^b(1-n_2^a-\frac{n_2^b}2)+n_2^b(1-n_1^a-\frac{n_1^b%
}2)]  \nonumber \\
&&-\frac t2\left[ 1-\frac{U-V+J^{(2)})}{2C}\right] \sum_{\sigma}
\sum_{i=1}^2[a_{i,\sigma }^{+}b_{\overline{i},\sigma }+b_{i,\sigma }^{+}a_{%
\overline{i},\sigma }],
\end{eqnarray}

\begin{equation}
H_D^{(7)}=E_{10}P_{10}+E_{11}P_{11}=2J^{(1)}[{S_1}^z\cdot {S_2}^z+\frac{%
n_1^an_2^a}4]
\end{equation}

\begin{eqnarray}
H_D^{(8)} &=&E_{12}P_{12}+E_{14}P_{14}=\frac{(t+U+J^{(1)}+J^{(2)})}%
4[n_1^an_2^b+n_2^an_1^b]  \nonumber \\
&&-\frac{(t+U+J^{(1)}+J^{(2)})}2\sum_{\sigma} [b_{1,\sigma }^{+}b_{2,\sigma
}+b_{2,\sigma }^{+}b_{1,\sigma }], \\
H_D^{(9)} &=&E_{13}P_{13}+E_{15}P_{15}=\frac{(-t+U+J^{(1)}+J^{(2)})}%
4[n_1^an_2^b+n_2^an_1^b]  \nonumber \\
&&+\frac{(-t+U+J^{(1)}+J^{(2)})}2\sum_{\sigma} [b_{1,\sigma }^{+}b_{2,\sigma
}+b_{2,\sigma }^{+}b_{1,\sigma }],
\end{eqnarray}

\begin{equation}
H_D^{(10)}=E_{16}P_{16}=\frac{(U+J^{(1)}+J^{(2)})}2n_1^bn_2^b,
\end{equation}

and $n_i^{a,b}=n_{i,\uparrow }^{a,b}+n_{i,\downarrow }^{a,b}$, $n_{i,\sigma
}^b=b_{i,\sigma }^{+}b_{i,\sigma }=n_{i,\sigma }n_{i,-\sigma }$ $(i=1,2)$, $%
d_{1(2)}=a_{1(2),\downarrow }b_{1(2),\uparrow }=c_{1(2),\downarrow
}c_{1(2),\uparrow }$, $\overline{i}=1$ if $i=2$ and $\overline{i}=2$ if $i=1$%
. The partial Hamiltonians ((9), (10)), ((11)-(15)), ((16), (17)) and (18)
belong to the subspaces of $N=1,2,3$ and $4$, respectively. The expressions
(9)-(18) are exact and when sum them up (see (8)) we obtain again the dimer
Hamiltonian in the form given by the expression (2), as it should be. The
decomposition (8) of the dimer Hamiltonian (2) into 10 different parts
(9)-(18) according to dimer energy levels possesses several, important
advantages. First, it explicitely visualizes the important intrinsic
two-site interactions, deeply hidden in the dimer Hamiltonian (2). Due to
the fact that the formulae (9)-(18) are exact the information about the
competitive interactions within the model for a dimer is complete. Second,
all of them are ascribed to each dimer energy level. It, however, means that
such interactions can be thermodynamically active only in the case when the
corresponding level will be occupied by electrons. Third, we can see that
the same types of interactions (but with different coupling constants)
belong to quite different energy levels. It also means that they do not need
to be thermodynamically active at the same time (it depends on the
occupation of the particular levels) and that the resulting properties of
the system depend on their competition. Fourth, the formulae (9)-(18)
visualize the important fact that with the increase of the averaged number
of electrons $n=<N>$ the system will pass through different phases,
depending on the result of the competition between different
thermodynamically activated two-sites interactions.

The most important two-site intrinsic interactions (leading to magnetism or
superconductivity), presented in the expressions (9)-(18) can be devided
into two classes. First of them belongs to magnetic interactions
(ferromagnetic, antiferromagnetic - it depends on the sign of the parametrs $%
J^{(1)}$, $J^{(2)}$,$U$ and $V$). Such interactions are present in the
formulae (11) and (15) and describe Ising type interactions with coupling
constants generated by $J^{(1)}$ and $J^{(2)}$. The formula (11) contains
also the transverse interaction between spins. The Heisenberg type magnetic
interactions can be seen in the first terms of the formulae (13) and (14),
generated by more complex coupling constants, expressed by the model
parameters $J^{(2)}$, $U$, $V$ and $t$. It is interesting to note that when $%
J^{(2)}=V=0$ the first term in the formulae (13) and (14) describes
ferromagnetic or antiferromagnetic interactions, similar to well-known $t-J$
model (see e.g. Refs [65]-[75], [89]) because the coefficient $\frac{2t^2}%
C\approx \frac{4t^2}{\mid U\mid }$ for large $\mid U\mid $ and is exactly
the same as in the case of a real lattice. The first terms in the
expressions (13) and (14) can also be considered as generalized $t-J$
interactions, valid in a more general case of the model (2). The coupling
constants at the first terms in (13) and (14) can be negative or positive.
Thus, they can describe competitive ferromagnetic or antiferromagnetic
Heisenberg interactions belonging to different dimer energy levels. In the
case of the conventional Hubbard model ($J^{(1)}=J^{(1)}=V=0$) the coupling
constants at the first terms in (13) and (14) are reduced to $-\frac{2t^2}C$
(13) and $\frac{2t^2}C$ (14) what means that the conventional Hubbard model
for a dimer describes ferromagnetic and antiferromagnetic Heisenberg
interactions which compete together but they belong to different dimer
energy levels. Let us note that the terms containing the products like $%
d_1^{+}d_2=b_{1,\uparrow }^{+}a_{1,\downarrow }^{+}a_{2,\downarrow
}b_{2,\uparrow }=c_{1,\uparrow }^{+}c_{1,\downarrow }^{+}c_{2,\downarrow
}c_{2,\uparrow }$ and $d_2^{+}d_1=b_{2,\uparrow }^{+}a_{2,\downarrow
}^{+}a_{1,\downarrow }b_{1,\uparrow }=c_{2,\uparrow }^{+}c_{2,\downarrow
}^{+}c_{1,\downarrow }c_{1,\uparrow }$ describe the hopping of the Cooper
pairs. Such terms, present in the second terms in (12), (13) and (14) are
typical for the KPPK superconductivity models (cf. e.g. Refs [59]-[64]) with
positive or negative coupling constants. Let us, however, note that this
type of interactions are also present within the model (2) also in the case $%
V=0$. The transversal products ${S_1}^{+}{S_2}^{-}$ (${S_1}^{-}{S_2}^{+}$)
are present in the formulae (11), (13) and (14). When using the second
quantization we obtain ${S_1^{+}S_2^{-}=-}c_{1,\uparrow }^{+}c_{2,\downarrow
}^{+}c_{1,\downarrow }c_{2,\uparrow }$ (${S_1^{-}S_2}^{+}=-c_{2,\uparrow
}^{+}c_{1,\downarrow }^{+}c_{2,\downarrow }c_{1,\uparrow }$) and these terms
describe intersite Cooper pairs (cf. e.g. Ref [33] and papers cited
therein). The application of the resonating valence bond approach (cf. Refs
[28], [29], [52]) allows also to treat the terms like $({\mathbf{%
\overrightarrow{S_1}\cdot \overrightarrow{S_2}}}-\frac{n_1^an_2^a}4$) in
(13) and (14) when introducing the pairing operator $f_{2,1}=$ $\frac 1{%
\sqrt{2}}(c_{2,\downarrow }c_{1,\uparrow }-c_{2,\uparrow }c_{1,\downarrow })$%
. For example, when we restrict ourselves to the subspace $n_1^a=n_2^a=1$ we
obtain $({\mathbf{\overrightarrow{S_1}\cdot \overrightarrow{S_2}}}-\frac
14)=-$ $f_{2,1}^{+}f_{2,1}$ and such terms lead to superconductivity (cf.
Ref. [29]). It is also interesting to note that even in the case when $%
J^{(1)}=$ $J^{(2)}=U=$ $V=0$ (see formulae (9)-(18)) the main competitive
interactions (ferromagnetic, antiferromagnetic and superconducting) are
always present when only the hopping parameter does not vanish ($t\neq 0$).

\section{ Conclusions}
Let us consider the case of a real lattice described by the Hamiltonian (1).
We assume that the number of lattice points $i(j)$ is equal to $N$ (even
number) and we decompose the lattice into a set of $M=\frac N2$ dimers
described by the dimer index $I,{\alpha} $ ($J,\beta $) where $I(J)=1,2,..,M$
and ${\alpha} (\beta )=1,2$. The Hamiltonian (1) can thus be replaced by the
equivalent form

\begin{equation}
\begin{array}{l}
\begin{array}{ll}
H= & \sum\limits_IH_{D,I}+\sum\limits_{I\neq J,{\alpha} ,\sigma }t_{I,{\alpha%
} ;J,{\alpha} }c_{I,{\alpha} ,\sigma }^{+}c_{J,{\alpha} ,\sigma
}+\sum\limits_{I\neq J,{\alpha} \neq \beta ,\sigma }t_{I,{\alpha} ;J,\beta
}c_{I,{\alpha} ,\sigma }^{+}c_{J,\beta ,\sigma }
\end{array}
\\ 
+\frac 12\sum\limits_{I\neq J,{\alpha} ,\sigma }J_{I,{\alpha} ;J,{\alpha}
}^{(1)}n_{I,{\alpha} ,\sigma }n_{J,{\alpha} ,\sigma }+\frac
12\sum\limits_{I\neq J,{\alpha} \neq \beta ,\sigma }J_{I,{\alpha} ;J,\beta
}^{(1)}n_{I,{\alpha} ,\sigma }n_{J,\beta ,\sigma } \\ 
+\frac 12\sum\limits_{I\neq J,{\alpha} ,\sigma }J_{I,{\alpha} ;J,{\alpha}
}^{(2)}n_{I,{\alpha} ,\sigma }n_{J,{\alpha} ,-\sigma }+\frac
12\sum\limits_{I\neq J,{\alpha} \neq \beta ,\sigma }J_{I,{\alpha} ;J,\beta
}^{(2)}n_{I,{\alpha} ,\sigma }n_{J,\beta ,-\sigma } \\ 
-\sum\limits_{I\neq J,{\alpha} }V_{I,{\alpha} ;J,{\alpha} }c_{I,{\alpha}
,\uparrow }^{+}c_{I,{\alpha} ,\downarrow }^{+}c_{J,{\alpha} ,\downarrow
}c_{J,{\alpha} ,\uparrow }-\sum\limits_{I\neq J,{\alpha} \neq \beta }V_{I,{%
\alpha} ;J,\beta }c_{I,{\alpha} ,\uparrow }^{+}c_{I,{\alpha} ,\downarrow
}^{+}c_{J,\beta ,\downarrow }c_{J,\beta ,\uparrow }
\end{array}
\end{equation}

where $H_{D,I}$ is the dimer Hamiltonian given by the expression (2) where
the lower dimer index $I$ in the operators should be introduced (as e.g. $%
c_{1,\sigma }\rightarrow c_{I,1,\sigma }$, etc.). The dimer Hamiltonian $%
H_{D,I}$ in (19) can, however, be diagonalized and replaced by the
expression (8) where again the dimer index $I$ in the operators appearing in
the expressions (9)-(18) should be introduced. The Hamiltonian (19),
equivalent to (1), describes now ''free'' dimers (first term in (19)) and
their interactions (the next 8 terms in (19)). It is then evident that the
Hamiltonian (19) contains explicitely all the competitive two-site
interactions present in the dimer Hamiltonian alone (see the formulae (8)
and (9)-(18)) but supplemented by the dimer interactions, represented by the
second and further terms in (19). The main difference between Hamiltonians
(1) and (19), both describing the same physics, lies in the fact that the
mentioned competitive magnetic and superconducting interactions are hidden
in the Hamiltonian (1) whereas in the Hamiltonian (19) they appear now in a
direct way. It, however, means that to find thermodynamical properties of
the model we should simultanously introduce order parameters in four
competitive channels (ferromagnetic, antiferromagnetic and superconducting
(intrasite and intersite pairing)) where a special attention should be paid
to the nontrivial fact that all competitive interactions are additionally
scattered between different energy levels and therefore their activities
have to be correlated with the occupation of these levels.

\end{document}